\documentclass[preprint]{aastex}
\pdfoutput=1

\newcommand   {\sgra}  {Sgr A$\textrm{*}$}

\renewcommand {\deg}   {\mbox{$^\circ$}}

\newcommand   {\arcs}  {\mbox{$^{\prime\prime}$}}

\newcommand   {\kms}   {\mbox{km\,s$^{-1}$}}
\renewcommand {\ga}    {\mbox{\rlap{\hbox{\lower5pt\hbox{$\sim$}}}\hbox{$>$}}}
\renewcommand {\la}    {\mbox{\rlap{\hbox{\lower5pt\hbox{$\sim$}}}\hbox{$<$}}}

\begin{document}



\def\kms {\hbox{km{\hskip0.1em}s$^{-1}$}} 
\def\msol{\hbox{$\hbox{M}_\odot$}}
\def\lsol{\hbox{$\hbox{L}_\odot$}}
\def\kms{km s$^{-1}$}
\def\Blos{B$_{\rm los}$}
\def\etal   {{\it et al.\ }}                     
\def\eg     {e.g.\ }
\def\psec   {$.\negthinspace^{s}$}
\def\pasec  {$.\negthinspace^{\prime\prime}$}
\def\pdeg   {$.\kern-.25em ^{^\circ}$}
\def\degree{\ifmmode{^\circ} \else{$^\circ$}\fi}
\def\ee #1 {\times 10^{#1}}          
\def\ut #1 #2 { \, \textrm{#1}^{#2}} 
\def\u #1 { \, \textrm{#1}}          
\def\nH {n_\mathrm{H}}

\def\ddeg   {\hbox{$.\!\!^\circ$}}              
\def\deg    {$^{\circ}$}                        
\def\le     {$\leq$}                            
\def\sec    {$^{\rm s}$}                        
\def\msol   {\hbox{M$_\odot$}}                  
\def\i      {\hbox{\it I}}                      
\def\v      {\hbox{\it V}}                      
\def\dasec  {\hbox{$.\!\!^{\prime\prime}$}}     
\def\asec   {$^{\prime\prime}$}                 
\def\dasec  {\hbox{$.\!\!^{\prime\prime}$}}     
\def\dsec   {\hbox{$.\!\!^{\rm s}$}}            
\def\min    {$^{\rm m}$}                        
\def\hour   {$^{\rm h}$}                        
\def\amin   {$^{\prime}$}                       
\def\lsol{\, \hbox{$\hbox{L}_\odot$}}
\def\sec    {$^{\rm s}$}                        
\def\xbar   {\hbox{$\overline{\rm x}$}}         

\shorttitle{}
\shortauthors{}

\title{The Discovery  of  Radio Stars  within 10\arcs\ of \sgra\,  at 7mm} 
\author{F. Yusef-Zadeh$^1$, 
D. A. Roberts$^1$, H. Bushouse$^2$, M.  Wardle$^3$,  W. Cotton$^4$, M. Royster$^1$,  \& G. van Moorsel$^5$}
\affil{$^1$Department of Physics and Astronomy and CIERA, Northwestern University, Evanston, IL 60208}
\affil{$^2$Space Telescope Science Institute, Baltimore, MD 21218}
\affil{$^3$Dept of Physics and Astronomy, Macquarie University, Sydney NSW 2109, Australia}
\affil{$^4$National Radio Astronomy Observatory,  Charlottesville, VA 22903}
\affil{$^5$National Radio Astronomy Observatory, Socorro, NM 87801} 

\begin{abstract} 
Very Large Array observations of the Galactic Center at 7 mm
have produced an image 
 of the   30\arcs\  surrounding  \sgra\, with a resolution of 
 $\sim$82$\times42$ milliarcseconds (mas). 
A comparison  with IR images taken simultaneously with the Very Large Telescope (VLT)
identifies  41 radio sources 
with  L-band (3.8 $\mu$m) stellar counterparts. 
The well-known young,   massive stars in the central \sgra\ cluster (e.g., 
IRS 16C, IRS 16NE, IRS 16SE2, IRS 16NW, IRS 16SW, AF, AFNW, IRS 34W and IRS 33E)
are detected with peak flux densities between 
$\sim$0.2 and 1.3 mJy.
The origin of the stellar radio emission  in the central cluster is discussed in terms of ionized 
stellar winds  with mass-loss rates in the range
$\sim 0.8-5\times10^{-5}$ \msol\ yr$^{-1}$.  
Radio emission from eight massive stars
is used as a tool for
registration between the radio and infrared frames with mas precision 
within a few arcseconds of \sgra. This  is similar to the established 
technique of aligning SiO masers and evolved stars except that radio stars 
lie within a few arcseconds of Sgr A*.
Our data  show a scatter of $\sim$6.5 mas 
in the positions of the eight radio sources that appear in both the L-band and 7 mm images.  
Lastly, we use  the radio and IR data to argue that members of IRS 13N 
are  Young Stellar Objects rather than 
dust clumps, supporting the hypothesis that   recent star formation has occurred near 
Sgr A*. 
\end{abstract}


\section{Introduction}

Most of the far-IR luminosity of the inner few pc of the Galactic Center (GC) region can be accounted for by the central 
cluster of emission-line stars distributed throughout the region (e.g., Gezari, Dwek \& Varosi 2003; Genzel, Eisenhauer \& 
Gillesseon 2010).  The central stellar cluster lies mainly within 1--10 acrcseconds\footnote{one arcsecond corresponds to 
0.039 pc at the GC distance of 8 kpc} of \sgra\ and consists of about one hundred young massive OB and WR stars distributed 
amongst several small clusters, such as IRS 16 and IRS 13, within disks made up of stars (Paumard \etal 2006; Lu \etal 2009).  
Near-IR observations imply strong ionized winds with velocities on the order of 700 \kms\ and a total mass-loss rate of 
$4\times10^{-3}$ \msol\ yr$^{-1}$ (Martins \etal 2007) arising from the collection of the so-called disk stars in the central 
cluster.  \sgra\ is thought to be fuelled by partial capture of the winds from massive stars in the central cluster.  The 
accretion rate was originally estimated from Br$\gamma$ line measurements, indicating mass-loss from individual cluster stars 
(Narayan \etal 1995; Coker \& Melia 1997; Goldston \etal 2005; Falcke \etal 2009). The variability of the accretion rate was 
predicted by numerical simulations and analytical calculations accounting for the motion of individual mass-losing stars 
orbiting \sgra, with the time averaged accretion rate was then estimated to be $\sim10^{-6}$ \msol\ yr$^{-1}$ (Quateart 2004; 
Cuadra \etal 2006, 2008).  Radio emission from stars can potentially provide independent determination of mass-loss rates in 
the form of ionized stellar winds, with implications for the accretion rate onto \sgra. In addition, these measurements can 
potentially identify the sites of the interaction of ionized stellar winds with gas near Sgr A*, constraining  physical 
parameters of the interstellar medium near the black hole.
Radio observations, 
however, have so far been unable to identify isolated massive stars in the central cluster. 

We present the highest resolution continuum image yet obtained of 
the inner 30\arcsec\ of the GC at 7 mm.  
Previous high resolution 1.3 and 3.5 cm observations of this region were used to
investigate the proper motion of  HII regions  
(Yusef-Zadeh, Biretta \& Roberts 1998; Zhao \& Goss 1988;  Zhao \etal 2009). 
Individual stellar sources could not be separated, however, from the ionized flow associated with orbiting gas 
(Zhao \etal 2009). Radio emission from ionized winds from young stars 
has already  been detected from 
the Arches and Quintuplet clusters projected within 12$'$ of Sgr A* (Lang et al. 2005).  
The high resolution images presented here show for the first time  radio emission from relatively isolated massive 
stars in the central stellar cluster centered on \sgra. We determine the mass loss rate from ionized stellar winds
to be quite similar to those in the Arches and Quintuplet clusters.  

The positions of compact radio  stars can also be used 
to precisely register 
the radio and IR frames within a few arcseconds of \sgra. 
Aligning radio and IR frames is  a
technique  that  locates the position  of the bright radio source
Sgr A*  in highly confused  IR images (Menten \etal 1997; Reid \etal 2003, 2007).
This well-established technique uses maser emission from IR identified evolved 
stars to carry out accurate astrometry.  
Here we use the same technique and 
show that  radio emission from hot stars can also be used 
for astrometric measurements.
Precision astrometry  using either radio stars or SiO masers  
has  the potential  
in the search for post-Newtonian 
deviations in the stellar orbit of the S2 star located within 0.2\arcs\ of \sgra. 
Lastly, a number of partially resolved radio sources have  been detected toward 
IRS 13E and IRS 13N at 7mm.  The sources in  IRS 13N are thought to be dust clumps 
with embedded stars or YSOs. Here, we use 7mm data to argue that
they are  not dust clumps and that they support  the YSO hypothesis.


\section{Observations and Data Reduction}

Detection of radio stars in the GC is not a trivial task.
This is due to the crowded environment, where \sgra\ and the extended ionized gas from the mini-spiral
dominate  the emission within the inner pc of \sgra. 
Due to the bright diffuse emission from the mini-spiral, 
it is virtually impossible to unambiguously identify individual stellar sources in low resolution radio continuum images. 
An additional difficulty with identifying radio sources  is
the quadratic increase of 
the scattering size of 
distant sources with  wavelength $\lambda^2$  (e.g., Bower \etal 2006). 
To overcome these difficulties, we have conducted  high resolution observations of the GC 
at 7 mm  with the Very Large 
Array (VLA)\footnote{Karl G. Jansky Very Large Array (VLA) of the National Radio Astronomy Observatory is a facility of the National
Science Foundation, operated under a cooperative agreement by Associated Universities, Inc.}. 
These observations also have the   advantage that  radio emission  
from the ionized winds of hot stars 
has  inverted spectrum (F$_{\nu}\propto\nu^{0.6}$), and thus the emission 
is stronger at shorter wavelengths (Panagia \& Felli 1975). 

The VLA was used in its A configuration
to carry out two pairs of
observations at 7mm, each separated by one day, on July 8--9, 2011 and August
31--September 1, 2011.  
These high resolution observations used two IFs, 128 MHz wide, centered on 
41.5 and 42.5 GHz. Each IF was composed of 64 channels, 2 MHz in width.
J1331+3030 and J1744$-$3116 were used as the primary flux and complex
gain calibrators, respectively. J1733$-$1304 was used to correct antenna
pointing errors once every hour and was observed every ten minutes to
calibrate complex gains and bandpass.  Individual data sets were
self-calibrated in phase and amplitudes to remove atmospheric phase
errors.  We fixed the average flux of \sgra\ to be 2.2 Jy in each
observation before 54 channels were combined and self-calibrated in
both phase and amplitude.  Due to the intrinsic variability of \sgra, flux
measurements taken from the combined radio image are uncertain by
$\sim10-15$\%.  This is because the average flux of Sgr A* was not the
same in each session.  The final image (Fig. 1) 
was constructed with a
resolution of $\sim82\times42$ mas with an rms
noise of 61 $\mu$Jy per beam and a dynamic range of $\sim3.5\times10^4$. The FWHM of the 
primary beam is 1$'$ at 7 mm.  The pair of observations is
separated by 53 days from each other, so the effective epoch of the
combined image is August 4, 2011. We employed background-subtracted
2D Gaussian fits using JMFIT in AIPS for measuring the properties of
individual sources in the final radio image.

Eight sources were used to register the radio and IR images: IRS 16C, IRS 16NE, IRS 16NW, IRS 16SE2, IRS 33E, IRS 34W, 
AF, and AFNW (see Table 1). An L band (3.8$\mu$m) image (Fig. 2) 
 of the GC region was taken with the European Southern Observatory Very Large Telescope (VLT) on July 7, 2011.
A color composite  image using radio and IR data is shown in Figure 1b. 
 The IR 
image, shown in green in  Figure 1b, 
is produced with the NaCo adaptive optics imager and covers a field of view of 
$\sim$43\arcsec. The eight radio sources are relatively isolated and are easily distinguished from extended ionized 
gas, as can be seen in Figure 1a. The registration of the radio and IR images was accomplished as follows. The RA and 
Dec positions of the eight sources were measured in the radio map produced from the July 8, 2011 data by fitting 
elliptical Gaussians to the sources. The formal errors in the centroids of these measurements are on the order of 6 
mas. Predicted x/y coordinates of the eight sources in the IR image were then computed by transforming the radio 
RA/Dec values into the IR image plane using the original World Coordinate System (WCS) definition of the IR image. We 
then measured the actual x/y positions of the eight sources in the IR image by fitting elliptical Gaussians with the 
IRAF/STSDAS task "n2gaussfit". The IR image is oversampled, with FWHM for point sources of $\sim$4.4 pixels, and a 
pixel scale of 0.0272 arcsec. The formal uncertainties of the centroids of the individual sources in the IR image are 
less than 1/100 of a pixel, or $<$1 mas. The predicted and measured x/y coordinates of the sources in the IR image 
were then used to compute a geometric transform between the radio and IR image spaces, using the IRAF task "geotran". 
The resulting transform was then applied to the IR image to assign a new World Coordinate System and ultimately to 
remap the IR image into the same WCS space as the radio image. The transform solution showed rms residuals to the fit 
of the eight source positions of $\sim$6.5 mas. 
We have measured positional errors of Sgr A* before the data is  
self-calibrated. The error in the absolute position of Sgr A* is  
estimated to be $\sim$2.8 mas.  
 The total error in absolute position of radio  stars                        
when cross correlated with  IR images is then estimated to  be $\sim$7.1 mas. 
The residual errors in the positions of the eight sources in the IR 
image space are listed in Table 1 (column 5). The residuals are consistent with the uncertainty of the RA/Dec 
measurements in the radio image ($\sim$6 mas) and also must include an unknown component of uncertainty due to 
residual distortions in the construction of the mosaicked IR L-band image.

\section{Results and Discussion}

\subsection{Radio Stars}

Figure 1a shows a grayscale version of the 7 mm radio image covering  
$\sim17''\times15''$  around \sgra.
The northern  and eastern  arm components  
of the extended mini-spiral components run to the NE and SE.
The kinematics of these extended features indicate that they orbit \sgra\ (e.g., 
Lacy \etal 1980; Roberts and Goss 1993; Zhao \etal 2009).
We detect a large number of compact radio sources, as labeled on 
Figure 1a,  within 10\arcs\ of \sgra. 
Table 1 gives the IR identified names, RA and Dec 
coordinates (increasing in RA), total positional errors,  residual astrometric errors, peak 
and integrated fluxes at 7 mm, the deconvolved beam angular sizes and the ID number for radio stars between 1 and 41.
The flux measurements have accounted for the primary beam attenuation.  
The fitted position of Sgr A* (ID 0) is also included in the Table. 
Several sources such as IRS 13E1, 13E2, 13E3, 13E5 and IRS 13N$\alpha$ to 13N$\gamma$ 
show radio counterparts. There  
are  also stellar sources in these clusters that 
are not identified by name but have radio counterparts, as listed in Table 1. 
To clarify   radio sources with stellar counterparts in Table 1, we 
added a to k at the end of each cluster member. 
The effective epoch of  these source detections taken from the  combined 7 mm image,   is August, 4, 2011. 
In table 1, the combined uncertainties in RA and Dec are listed in column 3 in mas for only nine  sources. 
The remaining sources were  either partially or fully resolved or too faint to get 
reasonable positional errors. 
Thus, the positional errors of these sources are not determined.

The nine  radio stars  
coincide  with  some of the most luminous sources in the GC and the Galaxy as a whole (Paumard \etal 2006).   
Most of them show P-Cygni profiles, thus presenting 
strong evidence for stellar winds  (Najarro \etal 1997; Martins \etal 2007). 
Although spectral index values of the radio stars  are not available, 
the   emission is most likely  thermal,  resulting from expanding ionized winds in the envelope 
of hot stars (Panagia \& Felli 1975). 
We expect an insignificant  contribution from  nonthermal emission at 7 mm  (Contreras \etal 1996).  
 
We determined the mass-loss rate of individual radio stars 
assuming  the standard model for a spherically-symmetric, homogeneous  wind of fully
ionized gas 
with $T=1\times 10^4~{\rm K}$\ and expanding with a constant terminal velocity
(Panagia \& Felli 1975). 
 We assumed twice solar metallicity, a mean molecular weight $\mu$=2 and adopted 
terminal velocities 
of IR stars, as given by Martins \etal (2007).  
Table 2 compares  the mass-loss rates of several radio and IR stars, with rates
determined from the radio measurements and from detailed model atmosphere
calculations (Martins \etal 2007). Integrated fluxes and terminal velocities are
also listed.
IRS 16NE and  IRS 16SW are binaries (Pfuhl \etal 2014), 
thus their model atmosphere calculations are not available. 
The mass-loss rates obtained using the radio and IR techniques 
agrees with each  other to within  a factor  of two with the exception of IRS 33E. 
The mass loss rate of this star derived from radio  measurements is 
a third of that from model atmosphere calculations given by Martin \etal (2007). 
This disagreement  may  reflect on the assumption that 
stellar winds are not clumpy and  there is 
no nonthermal emission from radio stars. 
We note that  mass-loss rate  estimates are similar to those of 
radio stars in the Arches and Quintuplet clusters (Lang \etal 2005). 



\subsection{Radio and IR Astrometry}

We selected  eight of the nine hot stars for astrometric measurements. 
To determine if the detected radio sources have IR stellar counterparts, we used an L-band (3.8$\mu$m) image which 
was taken  on July 7, 2001.
In order to register the L-band image with the VLA radio image,
we matched  eight stellar IR sources within  about 10\arcsec\ from \sgra\ that are detected in the 7 mm radio map.
Note that  there should not be any  offset in the position of stars due to proper motion 
because the radio and IR images  were  taken within one day of each other. 
The radio counterpart to IRS 16SW is 
in a confused region, so  was not used in astrometric analysis. 
Figure 2 shows a larger view of the region shown in Figure 1 at  3.8$\mu$m but with RA, Dec coordinates 
labeled.  
All 41 sources listed in Table 1  have IR counterparts.  
We have also  detected  radio star candidates 
with no IR counterparts. 


Previous studies used 
the stellar SiO masers located beyond  $\sim$7\arcs\ of \sgra\ to register the radio and IR frames 
(Menten \etal 1997; Reid \etal 2003, 2007). 
This technique achieves positional accuracy of individual masers to $\sim 1$ mas (Reid \etal 2007).  
Individual sources such as IRS 16NE 
show positional errors 
of   2.2 mas in the 7 mm radio image. 
Our approach of using radio stars to register the IR frame yields a statistical
accuracy of the registration $\sim$6.5 mas. The eight radio stars used in our registration
are  distributed within $\sim$10\arcs\ of \sgra. 
This registration has allowed us to cross-identify stellar sources in radio and infrared images. 


\subsection{Young Stellar Objects vs Dust Clumps}

In addition to the hot mass-losing  stars listed in Table 2, 
we detected  multiple radio sources coincident with IR-identified stars and stellar clusters: 
2 in IRS 1W, 4 in IRS 21, 6 in the IRS 13E cluster, 2 in IRS 5,  
and  11  in the  IRS 13N cluster
(Paumard \etal 2006; Lu \etal  2009; Maillard \etal 2004; 
Sch\"odel \etal 2005; Muzic \etal 2008;  
Perger \etal 2008; Fritz \etal 2010; Eckart \etal 2004, 2013).
Figure 3 shows 7 mm and L-band images of the IRS 13E and IRS 13N
clusters located 3.5$''$ SW of \sgra. 
IRS 13E consists of several early-type stars  (e.g., Fritz \etal 2010; Eckart \etal 2013). 
Stellar sources  IRS 16NE and IRS 16SW 
are binary systems and display  evidence of X-ray emission resulting from colliding winds
(Coker \etal 2002).

Unlike compact, hot stars, a number of radio sources are partially resolved, as shown in 
column 6 of Table 2. 
A  significant deconvolved size of hot stars could not be measured, 
thus they are likely to be unresolved. By 
contrast, sources in the IRS 13N cluster, such as IRS 13N$\beta$ (source 23 in Table 1) have typical deconvolved beam sizes of 
$77\times38$ mas. There is a debate as to whether some members of IRS 13E and IRS 13N clusters are dust clumps with embedded 
stars or Young Stellar Objects (YSOs; Muzic \etal 2008; Fritz \etal 2010; Eckart \etal 2013). 
There are two arguments against members of IRS 13N being  dust 
clumps. First,  IRS 13N has a flux density 1.34 Jy at 3.8\,$\mu$m (Viehmann \etal 2006), 
which yields a luminosity estimate $4\pi d^2\, \nu S_\nu = 6\times10^4\,\lsol$ after correcting for 2 magnitudes of 
extinction.  If IRS 13N is a dust clump with no embedded source, 
we estimate  a heating rate $\sim \pi (400\,\mathrm{AU})^2 * L_*/(4\pi\,(\mathrm{0.5\,pc})^2) \sim 
1200\,\lsol$, assuming a typical distance of 0.5 pc from the GC,  a source size of 800AU (0.1$''$) and 
UV  radiation filed of 
corresponding to a luminosity $L_* \approx 2\times10^7$\,\lsol\, from the GC 
(Serabyn \& Lacy 1985). 
The IR luminosity of IRS 13N exceeds the heating rate of IRS 13N 
 by an order of magnitude. External heating by the intense stellar UV 
radiation field in the inner parsec of the galaxy, is insufficient, so an internal source of heating is needed. This favors 
the suggestion that these sources are dusty stars. Eckart \etal (2013) have  made a number of arguments 
based on the motion of the IRS 16 cluster and the interstellar material surrounding the cluster that favor the  young age of 
the IRS 13N cluster. 
We  have also studied  
Spectral Energy Distribution (SED) fitting of IRS 13N  which is  consistent with  a grid of model  calculations of 
YSO candidates (Yusef-Zadeh et al. 2014, in 
preparation).


The second line of argument   is that the  radio emission 
is consistent with thermal bremsstrahlung from ionized gas that is  being photo evaporated from a disk by 
the UV radiation field from hot stars in the inner parsec of the Galaxy.  The measured radio flux, $\sim 2$\,mJy at 43\,GHz, 
implies a volume emission measure $\int n_e^2\,dV = 6\times10^{58}$\,cm$^{-3}$.  Assuming a homogeneous source of 
FWHM$\approx0.1''$  and $T_e=8\,000$\,K yields a source radius $\sim 400\,$AU and $n_e\approx 2.5\times10^5\,$cm$^{-3}$, with an 
optical depth $\sim0.02$.  The mass of ionized gas is then $M_i\approx 3\times10^{-4}\,  \msol$. 
Hot stars in the inner parsec of the Galaxy do not create sufficient numbers of ionizing photons to explain this mass of ionized gas.
 The total production rate is 
estimated to be $Q_\mathrm{LyC}\approx 2.5\times10^{50}$\,s$^{-1}$ (Genzel, Hollenbach \& Townes 1994), implying an incident 
ionizing photon flux $Q/(4\pi\,(\mathrm{1\,pc})^2) \approx 2\times10^{12}$\,s$^{-1}$\,cm$^{-2}$.  This is in rough agreement 
with the hydrogen recombination rate per unit area in the ionized gas, $\alpha^{(2)}n_e^2 r \approx 
1\times10^{12}$\,s$^{-1}$\,cm$^{-2}$.  Assuming that the ionized gas expands at the sound speed $c_s\approx 
15$\,km\,s$^{-1}$ for $T=8\,000$\,K, the mass-loss rate due to photo evaporation is $M_i/(r/c_s)\approx 
2.2\times10^{-6}\msol\, yr^{-1}$.  This must be replenished on the expansion time scale $r/c_s \approx 140$\,yr, implying the 
existence of a reservoir of neutral material, presumably a disk associated with the YSO.

\section{Summary}

We have detected a number of stellar sources within the inner 10$''$ of 
\sgra\ in  7 mm radio continuum images with  astrometric precision  of $\sim6.5$  mas.  
The radio  emission from GC stars is at a level consistent 
with ionized winds from hot massive stars with mass-loss rates  determined 
from  detailed modeling of the atmospheres.
(Najarro \etal 1997; Martins \etal 2007). 
We also detected several radio sources in IRS 13N and  IRS 13E suggesting
that thermal radio emission  from ionized gas  is being photo-evaporated from  disks 
of YSOs by UV radiation from hot stellar sources in the GC. 
This possibility  is consistent with 
the idea that ongoing star formation is taking place within 
0.5 pc of Sgr A* (Eckart \etal 2013; Yusef-Zadeh \etal 2013).

\acknowledgments
This work is dedicated to L.  Ozernoy who encouraged FYZ  
to search for ionized stellar winds at radio wavelengths in the mid 90's. 
We are grateful to S. Gillessen and the MPE group for providing us with  a VLT image 
of the Galactic center. 
We thank the referee for detailed comments. 
This work is partially supported by the grant AST-0807400 from the NSF.

\begin{figure}
\centering
\includegraphics[scale=0.55,angle=0]{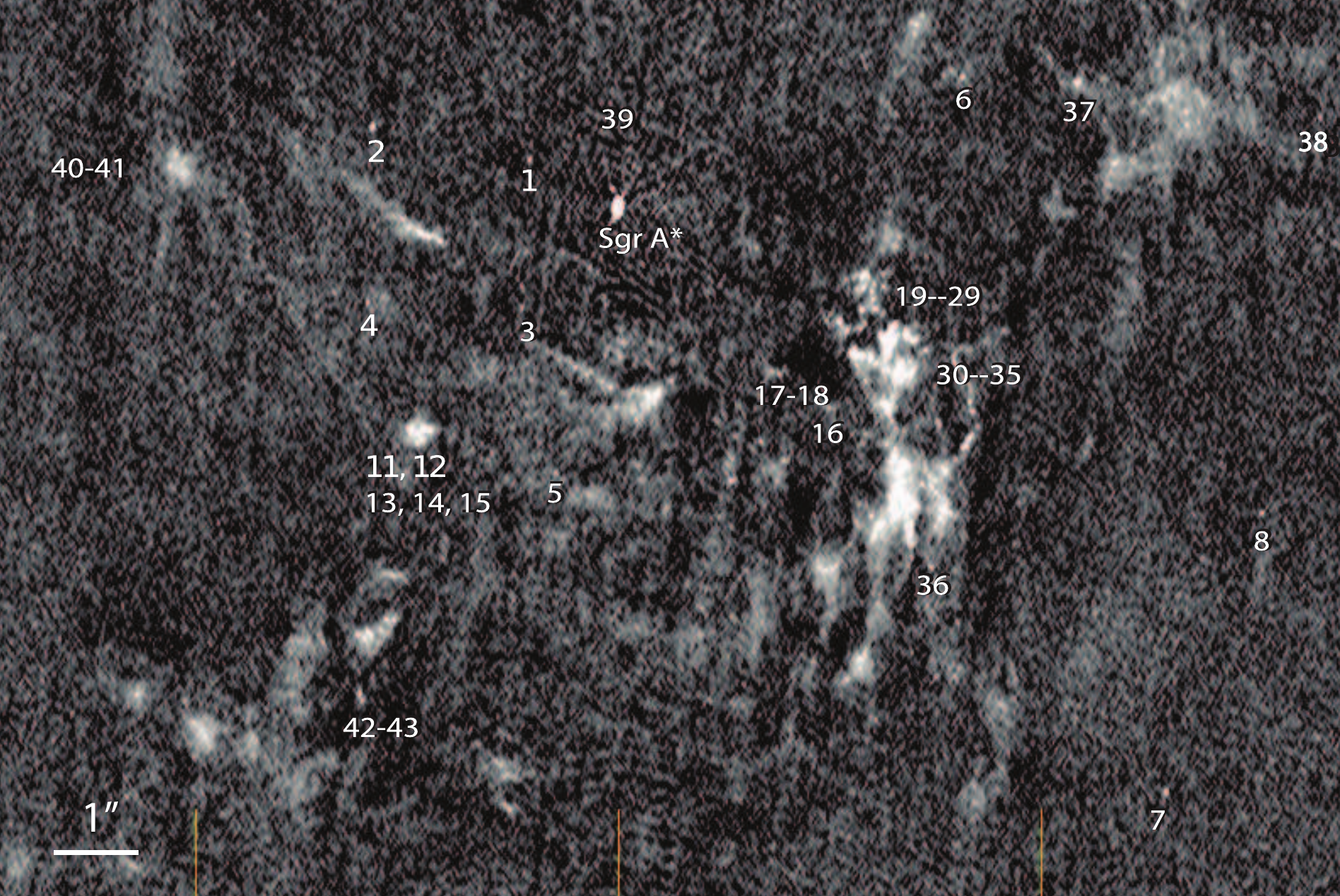}
\includegraphics[scale=0.55,angle=0]{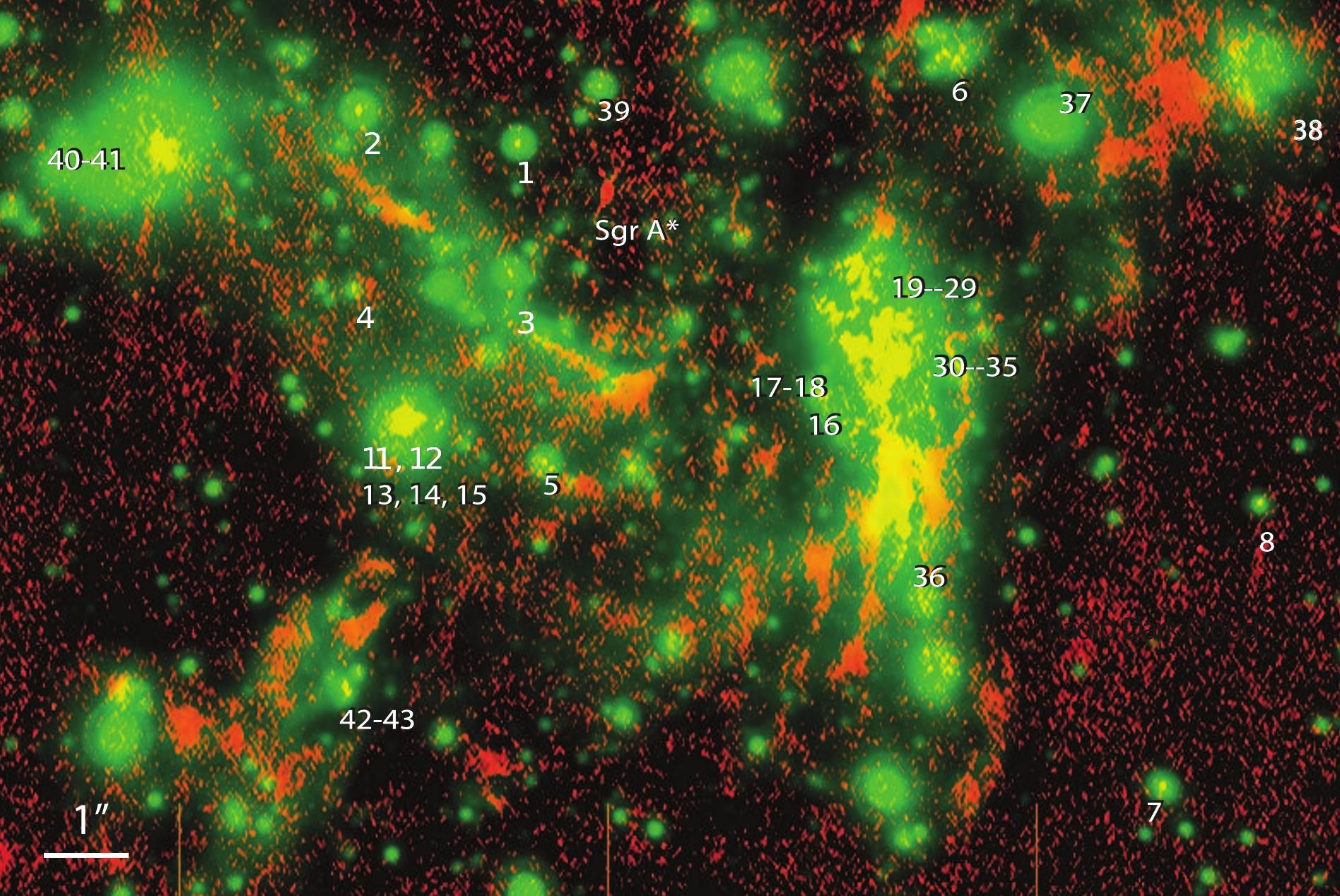}
\caption{{\it{(a)}} 
A  7 mm continuum image 
with a resolution of 82$\times42$\ mas (PA=$-5.5^\circ$). 
The  effective epoch  is August 4, 2011.  
{\it{ (b)}}
Same as (a) except a color composite image shows  L-band (3.8$\mu$m) image in green
and  a 7mm image in red.    
}
\end{figure}

\begin{figure}
\centering
\includegraphics[scale=0.45,angle=0]{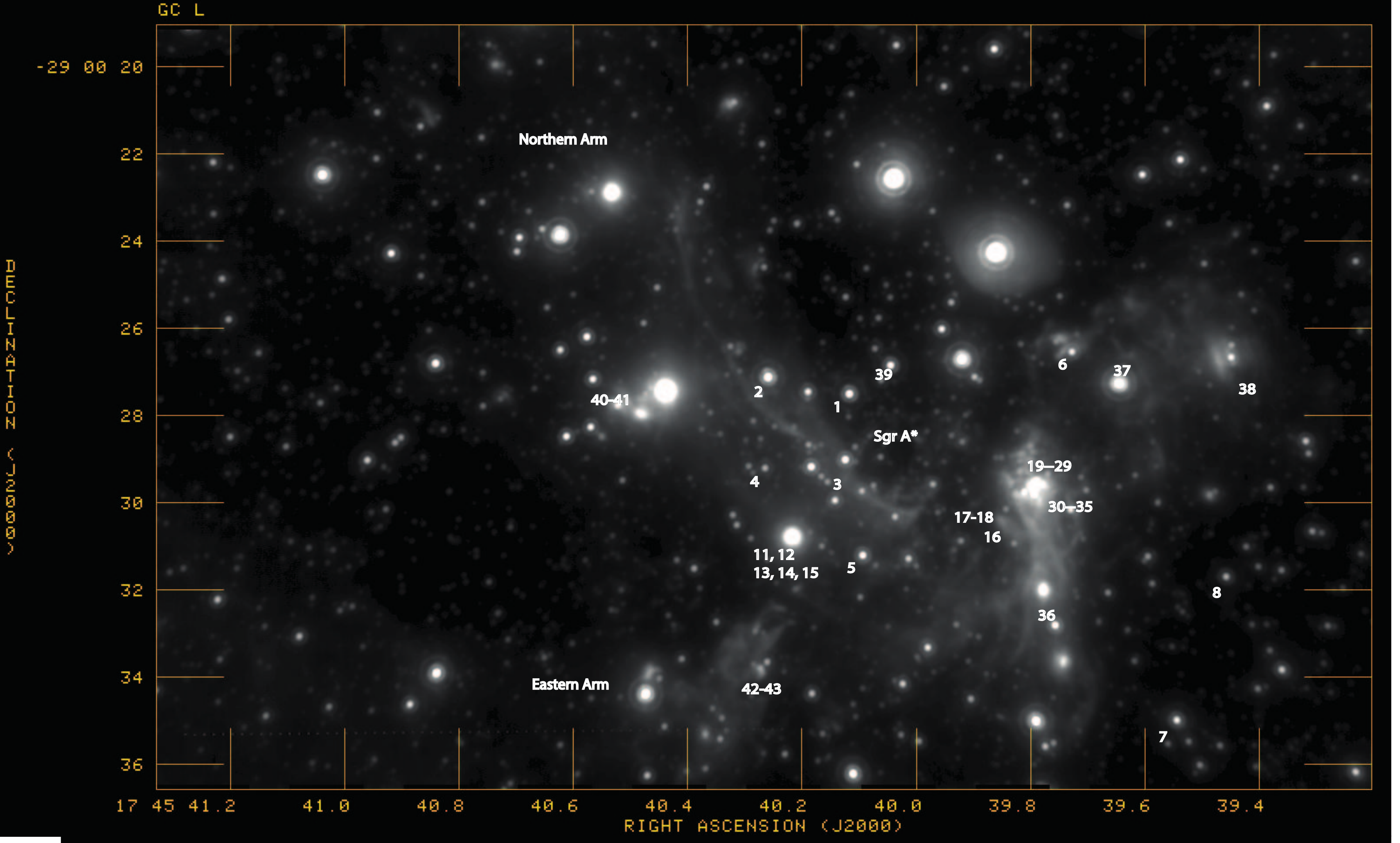}
\caption{
A  3.8$\mu$m image shows  a larger view than presented in Figure 1. 
All radio stars are  labeled  except sources 9 and 10  which lie 
outside the displayed region  (see Table 1). 
}
\end{figure}  


\begin{figure}
\centering
\includegraphics[scale=0.4,angle=0]{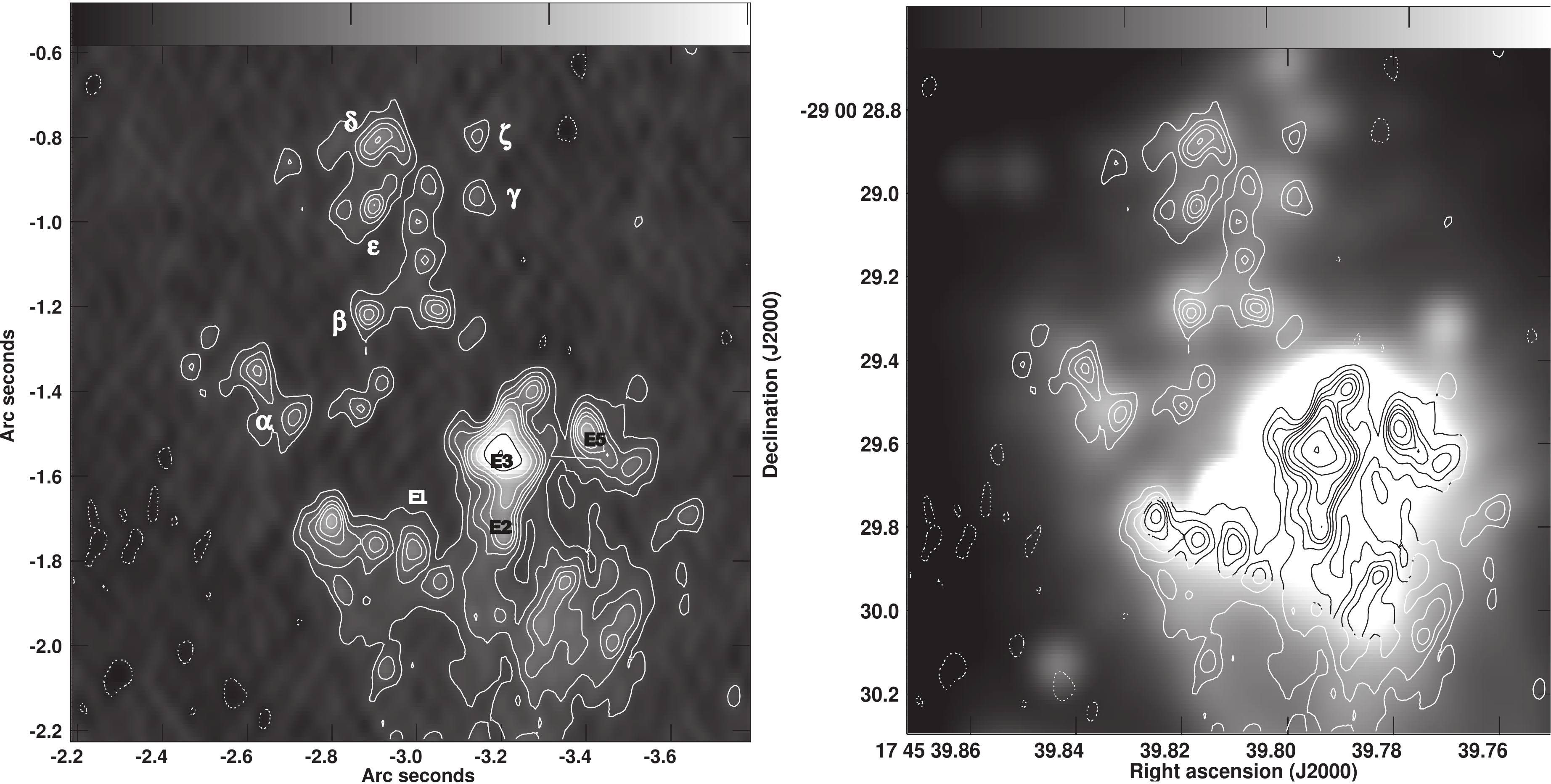}
\caption{
{\it{(a)}}
A grayscale   7 mm continuum image 
with an angular  resolution of 82$\times42$ mas (PA=$-5.5^\circ$).  
Contour levels  are set at 
(-1, 1, 2, 3, 4, 5, 7, 10, 15) $\times$ 0.22 mJy beam$^{-1}$. 
Prominent members of IRS 13E and IRS 13N are labeled. 
The coordinates are offsets from the position 
of Sgr A*.   
{\it{(b)}} Same as (a) except that contour levels are superimposed on 
an L-band (3.8$\mu$m) image and coordinates are in RA, Dec. 
}
\end{figure}  

\vfill\eject

\newcommand\refitem{\bibitem[]{}}
{}

\begin{deluxetable}{lcccccccc}
\tabletypesize{\scriptsize}
\tablecolumns{9}
\tablewidth{0pt}
\setlength{\tabcolsep}{0.04in}
\tablehead{
\colhead{IR Name}           & \colhead{RA (J2000)}      & \colhead{Dec (J2000)} & \colhead{Pos. Accuracy} & \colhead{Astrometric error} & $\theta_a\times\theta_b$ (PA)  & \colhead{Peak Int.}                    & \colhead{Int. Flux}       &  \colhead{ID} \\
                            & ($17^{\rm h} 45^{\rm m}$) & ($-29^{\circ} 00'$)   & (\sl{mas})\tablenotemark{\dag} & (\sl{mas}) & mas$\times$mas (deg) & (\sl{mJy beam$^{-1}$})\tablenotemark{\ddag} & (\sl{mJy})\tablenotemark{\ddag} &            \\ 
}
\startdata
Sgr A*                      & 40.0383                   & 28.069                & ---     & ---  & ---                      & 2,197.7                                     & 2,198.8                         &  0       \\
                            & 39.4051                   & 27.015                & ---     &---	 & ---                      & 0.49                                        & 0.37                    & 38      \\
AFNW\tablenotemark{\S}      & 39.4577                   & 31.686                & 3.4     &8.1	 & ---                      & 0.73                                        & 0.56                            &  8           \\
AF\tablenotemark{\S}        & 39.5440                   & 34.980                & 3.6     &18.1	 & ---                      & 0.84                                        & 0.91                            &  7           \\
                            & 39.6239                   & 26.602                & ---     &---	 & 53$\times15$ (94)                      & 0.84                                        & 1.36                            & 37           \\
IRS 34W\tablenotemark{\S}   & 39.7277                   & 26.535                & 3.3     &11.2	 & ---                      & 0.77                                        & 0.62                            &  6           \\
                            & 39.7756                   & 31.897                & ---     &---	 & 117$\times43$ (7)                      & 1.24                                        & 3.18                            & 36           \\
IRS 13E5-f                  & 39.7786                   & 29.568                & ---     &---	 &  ---                     & 1.60                                        & 2.06                            & 35           \\
IRS 13E3-d                  & 39.7941                   & 29.617                & ---     &---	 &  84$\times40$ (75)                    & 5.23                                        & 13.10                           & 33           \\
IRS 13E2-e                  & 39.7943                   & 29.784                & ---     &---	 & 132$\times66$ (5)                      & 1.56                                        & 1.67                            & 34           \\
IRS 13N$\gamma$-j           & 39.7987                   & 29.010                & ---     &---	 & ----                     & 0.57                                        & 0.41                            & 28           \\
IRS 13N-f                   & 39.8062                   & 29.274                & ---     &---	 & ----                      & 0.85                                        & 0.66                            & 24           \\
IRS 13N-k                   & 39.8075                   & 28.985                & ---     &---	 & 47$\times16$ (58)                      & 0.57                                        & 0.84                            & 29           \\
IRS 13N-g                   & 39.8083                   & 29.161                & ---     &---	 & ---                      & 0.52                                        & 0.61                            & 25           \\
IRS 13N-h                   & 39.8094                   & 29.070                & ---     &---	 & ---                      & 0.50                                        & 0.46                            & 26           \\
IRS 13E1-a                  & 39.8108                   & 29.811                & ---     &---	 & ---                      & 0.61                                        & 0.61                            & 30           \\
IRS 13N-d                   & 39.8162                   & 29.451                & ---     &---	 & ---                      & 0.65                                        & 0.98                            & 22           \\
IRS 13E-b                   & 39.8171                   & 29.831                & ---     &---	 & ---                      & 0.69                                        & 0.53                            & 31           \\
IRS 13N$\epsilon$-i         & 39.8173                   & 29.030                & ---     &---	 & ---                      & 0.97                        & 0.94                      & 27           \\
IRS 13N$\beta$-e            & 39.8181                   & 29.285                & ---     &---	 & $77\times38$ (175)                        & 0.89                                        & 0.77                            & 23           \\
IRS 13N-c                   & 39.8198                   & 29.505                & ---     &---	 & $76\times$41 (103)                      & 1.04                                        & 2.38                            & 21           \\
IRS 13E-c                   & 39.8250                   & 29.783                & ---     &---	 & $59\times32$ (62)                      & 1.64                                        & 3.01                            & 32           \\
X3                          & 39.8478                   & 30.445                & ---     &---	 & ---                      & 0.34                                        & 0.65                            & 16           \\
IRS 13N$\alpha$-a           & 39.8319                   & 29.530                & ---     &---	 & ---                      & 0.74                                        & 1.11                            & 19           \\
IRS 13N-b                   & 39.8382                   & 29.426                & ---     &---	 & ---                      & 1.05                                        & 2.58                            & 20           \\
                            & 39.8927                   & 30.045                & ---     &---	 & ---                      & 0.55                                        & 0.67                            & 17           \\
                            & 39.9004                   & 30.016                & ---     &---	 & ---                      & 0.36                                        & 0.48                            & 18           \\
IRS 16NW\tablenotemark{\S}  & 40.0451                   & 26.847                & 10.8     &28.8	 & ---                      & 0.39                                        & 0.27                            & 39           \\
IRS 33E\tablenotemark{\S}   & 40.0941                   & 31.213                & 5.5     &8.6	 & ---                      & 0.44                                        & 0.32                            &  5           \\
IRS 16C\tablenotemark{\S}   & 40.1177                   & 27.513                & 4.2     &8.0	 & ---                      & 0.71                                        & 0.68                            &  1           \\
IRS 16SW                    & 40.1245                   & 29.018                & 8.6     &---	 & ---                      & 0.42                                        & 0.47                            &  3           \\
IRS 21-a                    & 40.2095                   & 30.696                & ---     &---	 & ---                      & 1.12                                        & 2.35                            & 11           \\
IRS 21-b                    & 40.2183                   & 30.845                & ---     &---	 & ---                      & 0.45                                        & 0.56                            & 12           \\
IRS 21-c                    & 40.2194                   & 30.742                & ---     &---	 & ---                      & 0.75                                        & 0.37                            & 13           \\
                            & 40.2254                   & 30.718                & ---     &---	 & $117\times57$ (16)                      & 1.06                                        & 3.28                            & 15           \\
IRS 21-d                    & 40.2258                   & 30.705                & ---     &---	 & ---                      & 0.88                                        & 3.13                            & 14           \\
IRS 16NE\tablenotemark{\S}  & 40.2594                   & 27.126                & 2.2     &9.7	 & ---                      & 1.32                                        & 1.22                            &  2           \\
IRS 16SE2\tablenotemark{\S} & 40.2637                   & 29.200                & 4.3     &12.9	 & ---                      & 0.59                                        & 0.39                            &  4           \\
IRS 1W-a                    & 40.4370                   & 27.463                & ---     &---	 & $73\times45$ (12)                      & 0.38                                        & 0.76                            & 40           \\
IRS 1W-b                    & 40.4398                   & 27.597                & ---     &---	 & $19\times16$ (3)                      & 0.56                                        & 5.64                            & 41           \\
IRS 5-a                     & 40.6938                   & 18.268                & ---     &---	 & ---                      & 0.47                                        & 0.41                            &  9           \\
IRS 5-b                     & 40.6956                   & 18.179                & ---     &---	 & $15\times11$  (8)                      & 0.49                                        & 2.85                            & 10           \\
\enddata
\tablenotetext{\S}{radio stars used for astrometry}
\tablenotetext{\dag}{unless noted, positions are accurate to 25 mas}
\tablenotetext{\ddag}{15\% uncertainty in flux due to scaling to Sgr A*, which is a variable source}
\end{deluxetable}

\begin{deluxetable}{lcccc}
\tabletypesize{\scriptsize}
\tablecolumns{5}
\tablewidth{0pt}
\tablehead{IR Name 
& \colhead{$\dot{M}$ (Radio)}
& \colhead{$\dot{M}$ (IR)}
& \colhead{Integrated Flux} & \colhead{Terminal Velocity} \\
 & \sl{($10^{-5}$M$_{\odot}$ yr$^{-1}$)}  & \sl{($10^{-5}$M$_{\odot}$ yr$^{-1}$)}  & \sl{(mJy)}  & \sl{(km/s)}
}
\startdata
34W & 1.81  & 1.30 & 0.61 & 650\\
IRS 16NW & 0.90 & 1.1 & 0.27 & 600\\
IRS 16C & 1.75 & 1.1 & 0.68 & 650\\
33E & 0.76  & 2.20  & 0.32 & 450 \\
AF & 2.63 & 1.80  & 0.91 & 700 \\
AFNW & 2.10 & 3.2  & 0.56 & 800\\
IRS 16SE2 & 4.98  & 7.0  & 0.39 & 2500\\
\\
IRS 16NE  & 3.28 & ---  & 1.22 &  700 \\
IRS 16SW  & 1.58 & ---  & 0.47 &  700 \\
\enddata
\end{deluxetable}

\end{document}